# Prioritizing documentation effort: Can we do better?

Shiran Liu  Zhaoqiang Guo  Yanhui Li  Hongmin Lu  Lin Chen  Lei Xu  Yuming Zhou  Baowen Xu
State Key Laboratory for Novel Software Technology, Nanjing University

*Abstract*— **Code documentations are essential for software quality assurance, but due to time or economic pressures, code developers are often unable to write documents for all modules in a project. Recently, a supervised artificial neural network (ANN) approach is proposed to prioritize important modules for documentation effort. However, as a supervised approach, there is a need to use labeled training data to train the prediction model, which may not be easy to obtain in practice. Furthermore, it is unclear whether the ANN approach is generalizable, as it is only evaluated on several small data sets. In this paper, we propose an unsupervised approach based on PageRank to prioritize documentation effort. This approach identifies "important" modules only based on the dependence relationships between modules in a project. As a result, the PageRank approach does not need any training data to build the prediction model. In order to evaluate the effectiveness of the PageRank approach, we use six additional large data sets to conduct the experiments in addition to the same data sets collected from open-source projects as used in prior studies. The experimental results show that the PageRank approach is superior to the state-of-the-art ANN approach in prioritizing important modules for documentation effort. In particular, due to the simplicity and effectiveness, we advocate that the PageRank approach should be used as an easy-to-implement baseline in future research on documentation effort prioritization, and any new approach should be compared with it to demonstrate its effectiveness.**

## I. INTRODUCTION

CODE documentations play an important role in software quality assurance [1-6, 25-30]. High quality code documentation is helpful for program comprehension [31-33], test case generation [55-59], API Recommendation [62-67], fault localization [5, 68-74], bug detection [77-79], and program repair [84]. However, many studies [7-10] point out that, due to time and monetary pressures, developers are often unable to write documents for all modules in a project. As a result, they may write incomplete documentation or neglect writing documentation entirely.

Many recent studies [12-20] proposed automatic source code summarization to mitigate manual documentation effort, which automatically generates natural language summaries of software systems by extracting important information from code. However, McBurney et al.'s study [21] shows that the document quality of the state-of-the-art source code summarization approach is still lower than that of manually-written expert summaries in many aspects.

In order to reduce the workload of developers in documentation effort, McBurney et al. [22] recently explored the possibility of using static source code metrics and textual analysis of source code to automatically prioritize documentation effort. The purpose of this is to enable that precious effort can be paid on writing documents for important modules in a project. Based on three types of metrics, i.e. Static source Code Metrics (SCM), Textual Comparison Metrics (TCM), and Vector Space Model (VSM) [23], they built different supervised prediction models to classify modules (i.e. classes) in a software system into two categories: "important" and "non-important". The "important" modules refer to the modules that should be first documented by programmers. Based on an empirical study of five projects, they reported SCM was poor predictors whereas TCM and VSM were good predictors in documentation effort prioritization.

However, there are a number of limitations in McBurney et al.'s work. First, for a supervised approach, there is a need to use sufficient labeled training data to train the prediction model, which may be difficult to obtain in practice. This is especially true for a new type of projects or projects with little historical data collected. Second, their ANN approach not only has a high computation cost but also involves many parameters needed to be carefully tuned. This imposes substantial barriers to apply them in practice, especially for large projects. Third, as state in McBurney et al.'s study [22], it is unclear whether the ANN approach is generalizable, as the proposed ANN approach is only evaluated using several small data sets.

In this study, we attempt to address the following problem: can we do better in prioritizing documentation effort? From the viewpoint of practical use, we expect that there is a simple yet effective approach for practitioners. To this end, taking McBurney et al.'s work [22] as a starting point, we propose an unsupervised approach based on PageRank [24] to prioritize documentation effort. Unlike a supervised approach, the

This paragraph of the first footnote will contain the date on which you submitted your paper for review. It will also contain support information, including sponsor and financial support acknowledgment. For example, "This work was supported in part by the U.S. Department of Commerce under Grant BS123456."

All authors are with the State Key Laboratory for Novel Software Technology at Nanjing University, China.
E-mail: {shiranliu, naplus }@smail.nju.edu.cn; {yanhuili, hmlu, lchen, xlei, zhouyuming, bwxu}@nju.edu.cn.

PageRank approach does not need any training data or many different types of metrics to build the prediction model. Indeed, the PageRank approach identifies "important" modules only based on the dependence relationships between modules in the target software system. Furthermore, the PageRank approach has a low computation cost and is easy to implement, which is scalable to large software systems. In order to evaluate the effectiveness of the PageRank approach, on the one hand, we use the same data sets from open-source projects as used in McBurney et al.'s study [22] to conduct the experiments. On the other hand, we use six additional large data sets collected from open-source projects to conduct the experiments. Based on the nine data sets, we perform an extensive comparison between our PageRank approach and McBurney et al.'s ANN approach (the state-of-the-art approach). The experimental results show that the PageRank approach is superior to the state-of-the-art ANN approach in prioritizing important modules for documentation effort. As a result, we hence suggest that the PageRank approach should be used as an easy-to-implement baseline in future research on documentation effort prioritization and any new approach should be compared against to demonstrate its effectiveness.

The main contributions of this paper are as follows:
1. We propose a simple unsupervised approach based on PageRank to prioritize important modules for documentation effort.
2. By an extensive experiment, we demonstrate that the PageRank approach is superior to the state-of-the-art approach in prioritizing documentation effort.
3. We provide a replication package[1], including the data sets and scripts, to facilitate the external validation or extension of our work.

The rest of this paper is organized as follows. Section II introduces our research background. Section III presents the PageRank approach we proposed in detail. Section IV describes the experimental designs. Section V reports the experimental results in detail. Section VI discusses the influence factors for the PageRank approach. Section VII analyzes the threats to validity of our study. Section VIII concludes the paper and outlines the future work.

## II. BACKGROUND

In this section, we introduce our research background, including the role of code document in software quality assurance, code document effort prioritization, and the advantages and disadvantages of supervised vs. unsupervised prediction approaches.

### A. Role of code documentation in software quality assurance

Code documentation is a key component of software quality assurance. Good code documentation can greatly promote the understanding of software system, accelerate the process of learning and reusing code, increase developer productivity, simplify maintenance, and therefore improve the reliability of software [25-30]. In contrast, poor code documentation is one of the main reasons for the rapid deterioration of software system quality [28]. Therefore, code documentation is an irreplaceable necessity to enhance software reliability. To sum up, code documentation plays a fundamental role at least in the following areas of software quality assurance.

(1) Program Comprehension. Code documentation is an important aid for program comprehension during software development and maintenance [31-33]. It is a common strategy for programmers to understand project code starting with code documentation [3]. Combined code related design documentation can help participants achieve significantly better understanding than using only source code [34]. For example, many programmers wrote comments (a form of code documentation) to actively record the technical debt [53] in the code itself (that is, mark the test, improvement, and fix to be completed in the code with comments, also known as self-admitted technical debt [54]) to assist the subsequent software understanding and maintenance.

(2) Test case generation. Test case generation is among the most labor-intensive tasks in software testing. Because code documentation written by tabular expressions is precise, readable, and can clearly express the intended behavior of the code, such documentation documents are widely used in the test case generation [55-59], which makes evaluation of test results inexpensive and reliable. In general, they can be used to generate oracle [60] used to determine whether any test results (input and output pairs) meet the specification.

(3) API Recommendation. Application Programming Interfaces (APIs) are a means of code reuse. The goal of API recommendation techniques is to help developers perform programming tasks efficiently by selecting the required API from a large number of libraries with minimal learning costs. Clearly, API documentation (a typical type of code documentation, such as Javadoc[2]) is an important source of information for programmers to learn how to use API correctly [61]. In practice, API documentation is widely used in API recommendation [62-67]. By analyzing the similarity between words in API documentation and code context or natural language words in programming tasks, the accuracy of API recommendation can be enhanced.

(4) Bug detection. Bug detection techniques have been shown to improve software reliability by finding previously unknown bugs in mature software projects [75, 76]. Bug detection based on code comments is one of the most extensively studied bug detection techniques [77-79]. For a function or API, developers often write comments (natural language type or Javadoc type) to indicate the usage. An inconsistency between comments and body of a function indicates either a defect in the function or a fault in the comment that can mislead the function callers to

---

[1] http://github.com/sticeran/ProgramAndData

[2] https://www.oracle.com/technetwork/java/javase/documentation/index-jsp-135444.html

3introduce defects in their code. Bug detection based on code comments is to search for such inconsistencies to find bugs in software.

(5) Program repair. Automated program repair (APR) is a technique for automatically fixing bugs by generating patches that can make all failure test cases pass for a buggy program. Although APR has great potential to reduce bug fixing effort, the precision of most previous repair techniques is not high [80-83]. For a defect, often hundreds of plausible patches are generated, but only one or two are correct. In order to improve the precision of APR, code documentation have been applied in this field recently. For example, the literature [84] used Javadoc comments embedded in the source code to guide the selecting of patches. As a result, a relatively high precision (78.3%) is achieved, significantly higher than previous approaches [83, 85-87].

The above-mentioned works have a direct contribution on enhancing software reliability, and it can be seen that these works heavily depend on the code documentation. Undoubtedly, if there are high-quality code documentations, the effectiveness and efficiency of many quality assurance activities could be greatly improved.

*B. Code documentation effort prioritization*

Due to the fact that not all code always needs to be documented to a high quality level and the purpose of saving developers' precious time, McBurney et al. [22] recently proposed the concept of code documentation effort prioritization, which refers to "*Programmers must prioritize their documentation effort. The sections of code that are the most important for developers to understand should be documented first.*" In other words, code documentation effort prioritization refers to the priority division of the code modules itself (note: the granularity in [22] is at the class).

Since the term "code documentation effort prioritization" derives from the work of McBurney et al., we elaborate on their approach and work. They proposed a supervised artificial neural network (ANN) approach to automatically prioritizing important modules for documentation effort. They first collected the actual priority labels of classes (the labels are binary: "important" and "non-important"), then collected several types of code metrics, and finally compared supervised models trained by ANN. At the first step, to get the actual priority of classes, they collected user scores (i.e. expert experience) on five Java projects at the class-level and divided the classes in each project into two categories: "important" and "non-important". Users rated a class based on their perception of its importance. In their research, the top 25% of class files scored by users were marked as "important", while the rest were marked as "non-important". At the second step, to investigate the effectiveness of different types of code metrics in predicting documentation effort priorities, they collected three types of code metrics: **S**tatic source **C**ode **M**etrics (SCM), **T**extual **C**omparison **M**etrics (TCM), and **V**ector **S**pace **M**odel (VSM) metrics. Appendix A lists all metrics of SCM and TCM used in their study [22]. On a given project, they took the following measures to collect the corresponding metrics. For SCM, they analyzed source code to collect size metrics, complexity metrics, and object-oriented metrics (shown in appendix). For TCM, they compared source code and the corresponding project homepage to compute their textual similarity as well as binary variables indicating whether class/package names appeared at the same time (shown in appendix). For VSM, they collected tf/idf (term frequency/inverse document frequency) for each word in the source code. At the last step, they used artificial neural networks (ANN) [33] to build various supervised prediction models and performed a 10-fold cross validation for evaluating their effectiveness. Based on the experimental results, they found that VSM were the best predicators, while SCM were the worst predicators.

*C. Supervised vs. unsupervised approaches*

A supervised approach aims to learn a function, from labeled data (i.e. a number of instances with features and labels), to model the relationship between features and the corresponding labels observable in the data. Artificial neural network (ANN) is one of the most commonly used supervised techniques. In order to build a supervised model, a large amount of labeled data is often required. In contrast, an unsupervised approach aims to learn a distribution, from unlabeled data (i.e. a number of instances with features), to discover and present the interesting structure in the data.

Compared with an unsupervised approach, a supervised approach leverages prior knowledge in the training data and hence is often expected to have a higher prediction effectiveness. However, the disadvantages are three-fold. First, the labeled training data may incur a significant data collection cost or even are very difficult to obtain. Second, the training data and the testing data may have different distributions such that the prior knowledge from the training data may not be well applied to the testing data. Third, it is often computation-intensive to train a supervised model, especially for complex supervised modeling techniques such as ANN.

In our prior work, we found that, in defect prediction field, simple unsupervised models had a competitive or even superior prediction effectiveness compared with the existing supervised models in the literature [89]. Inspired by this work, we want to explore whether the same phenomena can be observed in documentation effort prioritization. Specifically, we want to explore whether PageRank [24] can be used to prioritize "important classes" for documentation effort. PageRank is a well-known algorithm that measures the importance of web pages. Given a set of hyperlinked set of web pages, the PageRank algorithm abstracts it as a graph, in which nodes correspond to web pages and edges correspond to hyperlinks. In this graph, a node is believed as important if it has many in-edges or has in-edges from important nodes. In other words, it takes into account both the number and quality of hyperlinks to a page when determining the importance of the page. In nature, PageRank is an unsupervised approach, as labeled training data are not needed. In the last decade, PageRank has been found many successful applications in software engineering [36-40, 45, 46], including fault localization [45] and crosscutting concerns mining [46]. If we find an unsupervised approach such as PageRank has a competitive or even superior prediction effectiveness to the-state-of-the-art supervised approach, this will lead to a large benefit for practitioners in documentation effort prioritization.

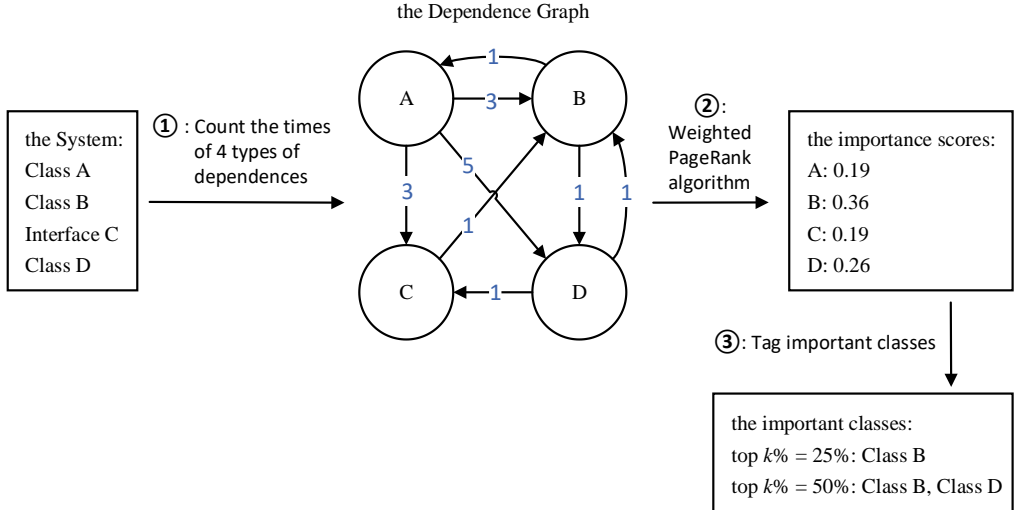

Fig. 1. An example object-oriented software system consisting of four modules

Fig. 2. The calculation process of our PageRank approach

## III. APPROACH

This section introduces our PageRank approach to documentation effort prioritization. First, at a high level, we outline the process of identifying important classes. Second, we give the definitions of dependence relationships used in our study. Third, we introduce how to compute the importance scores of classes by applying a weighted PageRank algorithm. Note that, in this study, a module can be a class or an interface in an object-oriented software system.

### A. Approach overview

For a given target project, our PageRank approach calculates the importance score of each module to complete the documentation effort prioritization. Specifically, the importance score calculation consists of the following three steps. First, build an inter-module dependence graph by analyzing the dependences between modules in the target software. Second, calculate the importance score of each module according to an extended PageRank algorithm. Third, sort the importance scores in descending order and classify modules whose importance scores exceed a threshold (e.g. top 5%) as important modules. These important modules are the ones that should be documented first.

We next use an example shown in Fig. 1 to illustrate our approach. As can be seen, the example target software system consists of four modules: three classes (i.e. A, B, and D) and one interface (i.e. C). In Fig. 1, we use annotations (dependence type and modules that are depended on) to explicitly describe the dependence relationships of each class with other classes. For example, for the second line of code in class A: "`public D d = new D ()`", we can see that the type of attribute d of class A is class D. Therefore, class A and class D have a Class-Attribute dependence relationship (i.e. CA, class A depends on class `D`). In order to describe this fact, we add a comment "`//CA: D`" to the end of this line. In our study, we take into account the following four types of dependences among modules:

(1) Class-Inheritance dependence (CI): One class (interface) is a subclass (interface) of another class; one class implements an interface in which the implemented interface is regarded as the parent class and the class implementing the interface is regarded as the subclass.
(2) Class-Attribute dependence (CA): The type of a class's (interface's) attribute is another class (interface).
(3) Class-Method dependence (CM): The parameter type or return value type of a class's method is another class (interface); the parameter type of an interface's method is another class (interface).
(4) Method-Method dependence (MM): A method in one class calls another class's method.

As stated in [36], the above four types of dependences are the main inter-module dependences in an object-oriented software system. Note that these four types of dependences can be efficiently obtained by a static analysis.

Fig. 2 shows the calculation process of our PageRank approach. As can be seen, a dependence graph is constructed based on the static analysis of source code. The dependence graph is a directed weighted graph. In this graph, each node represents a module, the direction of each edge represents the dependence direction, and the weight on each edge represents the total times of a module depends on another module. For example, the total times of class `A` depending on class `B` is 3 (one CI: `B` and two CM: `B`). Therefore, the weight on the edge <`A`, `B`> is 3. Based on the dependence graph, we use a weighted PageRank algorithm (see Section III C for detail) to calculate the importance score of each module. The higher the importance score is, the more important the module is. According to the score ranking, the top k% modules are regarded as important modules. These important modules are the ones that developers should document first. For example, if k% = 25%, only B is in the top 25%. Therefore, B is recognized as an important module for documentation effort.

### B. Generating edge-weighted module dependence graph

Given a target software system, by static analysis, we can easily extract the following dependence relationships among modules: Class-Inheritance dependence (CI), Class-Attribute dependence (CA), Class-Method dependence (CM), and Method-Method dependence (MM). Based on these dependences, we generate a directed weighted graph to depict the relationships among modules.

For the simplicity of presentation, in the following, we formalize the types of dependences and dependence graph. For a given object-oriented software system $S$, let $C(S)$ be the set of all modules (non-inner classes or interfaces) in $S$. Furthermore, let $|C(S)|$ be the number of elements in $C(S)$.

**Definition 1. Inter-module dependence**. Given a software system $S$, for $u, v \in C(S)$ and $u \neq v$, we have the following definitions:

(1) $CI(u, v)$ denotes the number of inheritance dependence between $u$ and $v$. If $v$ is the parent of $u$, the value is 1, otherwise it is 0.
(2) $CA(u, v)$ denotes the number of class-attribute dependences between $u$ and $v$, which is equal to the number of attributes' type of $v$ in $u$.
(3) $CM(u, v)$ denotes the number of class-method dependences between $u$ and $v$, which is equal to the number of methods in $u$ with $v$ as the parameter type or return value type.
(4) $MM(u, v)$ denotes the number of method-method dependences between $u$ and $v$, which is equal to the number of times of methods in $v$ is called in $u$.

**Definition 2. Edge-weighted module dependence graph**. Let $(u, v)$ denote one edge and $W(u, v)$ denote the weight of one edge. For a software system $S$, the corresponding dependence graph is defined as $G_S = (N, E, W)$, where

- $N = C(S)$
- $E = \{(u, v) \mid u, v \in N \land u \neq v \land W(u, v) > 0\}$
- $W = \{W(u, v) \mid W(u, v) = CI(u, v) + CA(u, v) + CM(u, v) + MM(u, v)\}$.

**Definition 3. In-Edge and Out-Edge**. For any node $u$ in a dependence graph $G_S = (N, E, W)$, its In-Edge and Out-Edge are defined as:

- In-Edge$_u$ = $\{(v, u) \mid v \in N \land v \neq u \land W(v, u) > 0\}$
- Out-Edge$_u$ = $\{(u, v) \mid v \in N \land u \neq v \land W(u, v) > 0\}$

### C. Deriving importance score by improved PageRank-VOL

Consistent with the original PageRank algorithm's criterion [24] for identifying the importance of web pages, we identify important modules according to the following criterion: a module is important if and only if many important modules depend on it. This criterion recursively defines whether a module is important depends on two factors: (1) the more modules depend on it, the more important it is; (2) the more important modules depend on it, the more important it is. Inconsistent with the original PageRank algorithm, we further take into account the third factor: the strength of dependence among modules. In other words, the higher strengths depend on it, the more important it is.

To this end, we use the improved PageRank based on Visits Of Links (PageRank-VOL) proposed by Kumar et al. [41] to calculate importance scores of modules. The reason why we choose PageRank-VOL is that the idea of this method is consistent with the idea of edge-weighted dependence graph proposed by us. The PageRank-VOL has considered the number of outgoing and inbound links between web pages and it assigns more rank value to the outgoing links which is most visited by users. In this manner, a page rank value is calculated based on visits of inbound links. Indeed, PageRank-VOL is a special case of edge-weighted PageRank [47].

The original calculation formula of PageRank-VOL is:

$$PR(u) = d \sum_{v \in B(u)} \frac{L_u PR(v)}{TL(v)} + (1 - d) \quad (1)$$

Based on the above equation, we change the "(1-$d$)" part to "(1-$d$)/$m$". Consequently, we have the PageRank-IVOL equation:

$$PR(u) = d \sum_{v \in B(u)} \frac{L_u PR(v)}{TL(v)} + \frac{(1-d)}{m} \quad (2)$$

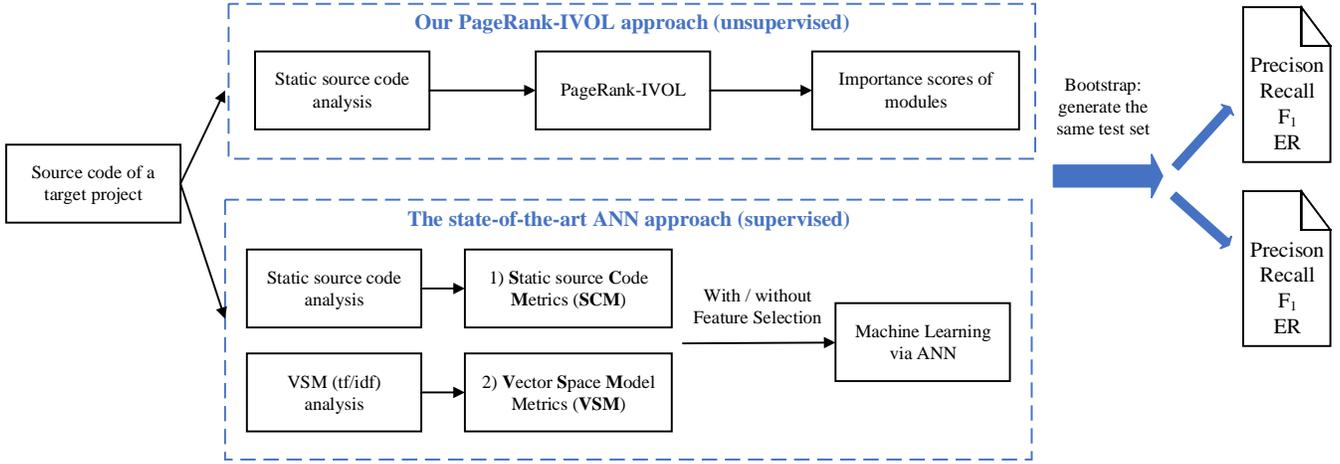

Fig. 3. Overall research framework in this paper

The objective of this goal is to ensure that the sum of the importance score over all the nodes in an edge-weighted module dependence graph equals to 1. The notations in our importance score equation are:

- $d$ is a dampening factor and it is generally taken 0.85.
- $m$ is the total number of nodes (i.e. modules).
- $u$ and $v$ ($u \neq v$) represent one node respectively.
- B($u$) is the set of nodes that point to $u$.
- PR($u$) and PR($v$) are important scores of node $u$ and $v$ respectively,
- $L_u$ is the weight of the *In-Edge$_u$* which points to node $u$ from $v$
- $TL(v)$ is the sum of weight of all *Out-Edge$_v$* which point to other nodes from $v$

In our implementation, we use the iterative method to solve PageRank-IVOL. The termination condition is that the error is less than $10^{-7}$ or the number of iterations reaches 100.

To explain the working of PageRank-IVOL, let us take the dependence graph in Fig. 2 as an example. The PageRank for nodes A, B, C, and D are calculated as follows:

PR(A) = d(PR(B)*1/2) + (1-d)/4
PR(B) = d(PR(A)*3/11 + PR(C) + PR(D)*1/2) + (1-d)/4
PR(C) = d(PR(A)*3/11 + PR(D)*1/2) + (1-d)/4
PR(D) = d(PR(A)*5/11 + PR(B)*1/2) + (1-d)/4

Let $P_0$ be the initial column vector whose sum of column elements is 1, where each element represents the initial importance score for each node. Furthermore, let $e$ be a column vector in which each element is 1. $M$ is the transition probability matrix. Then, we have:

$$P_0 = \begin{bmatrix} PR_0(A) \\ PR_0(B) \\ PR_0(C) \\ PR_0(D) \end{bmatrix}$$

$$P_1 = d \cdot M \cdot P_0 + \frac{1-d}{4} \cdot e$$

$$P_1 = d \cdot \begin{bmatrix} 0 & \frac{1}{2} & 0 & 0 \\ \frac{3}{11} & 0 & 1 & \frac{1}{2} \\ \frac{3}{11} & 0 & 0 & \frac{1}{2} \\ \frac{5}{11} & \frac{1}{2} & 0 & 0 \end{bmatrix} \cdot P_0 + \frac{1-d}{4} \cdot \begin{bmatrix} 1 \\ 1 \\ 1 \\ 1 \end{bmatrix}$$

……

$$P_n = d \cdot M \cdot P_{n-1} + \frac{1-d}{4} \cdot e$$

Generally speaking, the initial important scores are the same, so $PR_0(A) = PR_0(B) = PR_0(C) = PR_0(D) = 0.25$. The iteration of the above equation is continued until the termination condition is reached (e.g., the error is less than $10^{-7}$ or the number of iterations reaches 100). As a result: PR(A) = 0.19, PR(B) = 0.36, PR(C) = 0.19, and PR(D) = 0.26.

IV. EXPERIMENTAL DESIGNS

In this section, we introduce the experimental design to investigate the effectiveness of our proposed PageRank approach in prioritizing code documentation effort, including the research framework, baseline approaches, data sets, performance indicators, and evaluation method.

*A. Research framework*

Our motivation is to investigate whether there is a simple yet effective approach to documentation effort prioritization compared with the state-of-the-art ANN approach proposed in [22]. To the best of our knowledge, the ANN approach proposed in [22] is the first research to suggest an approach for automatically prioritizing code documentation. However, in [22], the state-of-the-art ANN approach is not compared against any baseline approaches. Consequently, it is unclear whether it is worth applying such a complex supervised approach in practice. In order to fill this gap, we choose to apply an improved PageRank to documentation effort prioritization, which is easy to implement. Our opinion is that, if the PageRank approach has a competitive effectiveness compared with the state-of-the-art ANN approach, it will more practical for practitioners to use a simple approach rather than a complex approach. This is especially true when taking into account their implementation cost and scalability to large projects.

TABEL I
(a) Old data sets

| Project | Version | The total number of classes | The total number of classes used in experiment [1] | Important classes (top 25%) | Source of Category labels |
|---|---|---|---|---|---|
| NanoXML | nanoxml-2.2.1 | 28 | 21 | 6 | |
| JExcelAPI | jexcelapi-2.6.12 | 458 | 50 | 11 | [22] |
| JGraphT | jgrapht-0.9.1 | 194 | 176 | 47 | |

1. In the Old data sets, the number of classes used in the experiment is equal to the number of classes scored by users.

(b) New data sets

| Project | Version | The total number of classes | The total number of classes used in experiment | The number of important classes | Source of Category labels |
|---|---|---|---|---|---|
| Ant | ant-1.6.1 | 664 | 664 | 8 | |
| ArgoUML | argouml-0.9.5 | 823 | 823 | 12 | |
| jEdit | jedit-5.1.0 | 539 | 539 | 7 | [39] |
| JHotDraw | jhotdraw-6.0b.1 | 498 | 498 | 9 | |
| JMeter | jmeter-2.0.1 | 224 | 224 | 13 | |
| Wro4j | wro4j-1.6.3 | 357 | 357 | 12 | |

Our overall research framework is shown in Fig. 3. On the one hand, for a given target project, we use an improved PageRank to compute module importance score based on the inter-module dependence graph generated by static source code analysis. On the other hand, we strictly follow the description in [22] to implement the ANN approach. Specifically, we implement two types of ANN models: the SCM-ANN model built with static source code metrics and the VSM-ANN model built with vector space model metrics. After that, the same test data set is used to compare the predictive effectiveness of the PageRank-IVOL approach and the ANN approach, including F1, recall, and precision.

### B. Baseline approaches

We use McBurney et al.'s metrics-based ANN approaches [22] as the baseline approaches to evaluate the effectiveness of our PageRank approach. In [22], McBurney et al. used the following four group metrics to build the ANN prediction models: (1) only the static source code metrics (SCM); (2) only the text comparison metrics (TCM); (3) only the VSM metrics; and (4) SCM + TCM. In our study, we use their SCM-based ANN model and the VSM-based ANN model as the baselines to investigate the effectiveness of our PageRank approaches. Note that we do not include the ANN models built with TCM or SCM + TCM as the baselines. The reasons are three-fold. First, they were inferior to the VSM-based ANN model according to the experimental results in McBurney et al.'s study. Second, it is not uncommon that the homepage of a project does not evolve with the project versions consistently, thus leading to the information mismatch between the homepage and source code. That is to say, we cannot guarantee that the content of the current project homepage corresponds to the version of the project we use. Third, what is more important, they are not applicable if a project does not have a corresponding home page. In contrast, both the SCM-based and VSM-based ANN model are based on only the source code of a project and hence are easier to be applied in practice.

We use the same ANN classifier and settings as used in McBurney et al.'s study [22] to build the ANN models. Specifically, the used classifier is the multilayer neural network library in the Neuroph[3] framework with the following settings: two hidden layers, the maximum number of iterations of 1000 runs, the learning rate is 0.25, the number of hidden nodes is 10, and a feed-forward fully connected architecture. Considering the large number of metrics in SCM and VSM, we generate four types of supervised ANN models: SCM model (static code metrics without feature selection), VSM model (VSM metrics without feature selection), SCM_FS model (static code metrics with feature selection), and VSM_FS model (VSM metrics with feature selection). In our study, we use CFS [42] (the Correlation-based Feature Selection) as the feature selection method to build the SCM_FS and VSM_FS models, by which the number of features reserved can be automatically determined. In McBurney et al.'s study [22], they did not apply feature selection to the VSM tf/idf metrics, i.e. the VSM_FS model was not considered. However, for a project, the VSM metrics often consists of thousands of metrics. This means that the number of features is far larger than the number of instances in a data set. Given this satiation, it will be very likely to produce an overfitted model. In this case, the resulting model may not generalize well to unseen data. Therefore, in our study, we take into account the VSM_FS model.

### C. Data sets

Our data sets are divided into two parts: Old data sets and New data sets, as shown in Table I(a) and I(b). The old data sets are from McBurney et al.'s study [22], while the new data sets are from Sora's study [39]. Specifically, the old data sets consist of three open-source Java API libraries (i.e. NanoXML 2.2.1, JExcelAPI 2.6.12, and JGraphT 0.9.1), while the new data sets of six open-source Java projects (i.e. Ant 1.6.1, ArgoUML 0.9.5, jEdit 5.1.0, JHotDraw 6.0b.1, JMeter 2.0.1, and Wro4j 1.6.3). In each data set, one instance corresponds to a class and consists of two parts: a number of metrics (i.e. SCM and VSM) and a 8label indicating whether it is important or not.

---

[3] http://neuroph.sourceforge.net/

In our study, we first download the source code for the projects used in McBurney et al.'s and Sora's studies. Then, we take the following measures to collect the data.

- Label data collection. For each project in the old data sets, McBurney et al. [22] published the importance of the classes used in their experiments, including class name and the corresponding subjective score evaluated by graduate students based on their perception of importance. Based on the importance score, they marked the top 25% classes as "important" and the rest as "non-important". For each project in the new data sets, Sora [39] published the names of important classes identified via the information from diverse sources, including the project tutorial, design description, development documents, and code analysis. As a result, for each class in the old and new data sets, we can easily obtain the corresponding label (i.e. "important" or "non-important").
- Metric data collection. In order to collect SCM, for each project, we first used a commercial tool Understand[4] to generate the corresponding udb database by parsing its source code. The udb database stored the entities and their relationships in the project. Then, we develop a Perl script to collect SCM based on the udb database. In order to collect the VSM metrics, similar to [22], we first developed a Python script to remove all special characters and numbers. Then, we generated the VSM metrics via Weka's StringToWordVector filter.

From Table I(a), we can see that, for each project in the old data sets, not all the classes are used in McBurney et al.'s study. On a given project, McBurney et al. invited three graduate students to score classes according to their importance for document with respect to program comprehension. Before scoring, each participant was asked to do a programming task involving the project (at most 70 minutes). Then, the participant used the entire remainder of the 90-minute time limit to score the importance to document each method within a class (the order of the methods was randomized). After, each class was assigned "the average of the average scores for each scored method in the class" [22]. Due to the 90-minute time limit, not all the classes in a project were scored by the participants. Consequently, the identified important classes, obtained from the scored (randomly chosen) classes in a project, may not be the real important classes for the whole project. In this sense, the old data sets are indeed problematic and their experimental conclusion on the effectiveness of their ANN approach may be unreliable. This problem is especially true for NanoXML 2.2.1 and JExcelAPI 2.6.12: 75% classes in NanoXML 2.2.1 were scored while only 11% classes were scored in JExcelAPI 2.6.12. Nonetheless, in our study, we keep the old data sets for our experiments. The purpose of this is to enable a direct comparison of our PageRank approach with the ANN approach reported in their study.

From Table I(b), we can see that, for each project in the new data sets, all the classes in the whole project were involved to identify the important classes [39]. Furthermore, diverse information sources, mainly provided by the original developers of the project, are used to label important classes. For Ant, the important classes were provided in its project tutorial and development document. For ArgoUML, the important classes were provided in its detailed architectural descriptions. For jEdit, the important classes were pointed out by its development document. For JHotDraw, the important classes were provided in its development document and complemented by code analysis. For JMeter, the important classes were mentioned in its design document. For Wro4j, the important classes were provided in its design overview. As can be seen, compared with the old data sets, the new data sets have the following advantages: (1) the important classes are provided by the original developers rather than the users of a project; (2) the important classes are identified on the basis of the whole project rather than a part of the project; and (3) the number of instances for each project is large enough to draw meaningful conclusions. In this sense, the new data sets are quality data sets for evaluating the effectiveness of our PageRank approach as well as the state-of-the-art ANN approach.

|  |  | Actual label | |
|---|---|---|---|
|  |  | Important modules | Non-important modules |
| Predicted label | Important modules | TP | FP |
|  | Non-important classes | FN | TN |

Fig. 4. The confusion matrix

### D. Performance indicator

According to [22], in nature, documentation effort prioritization is a classification problem: the modules (i.e. classes and interfaces in our study) are classified into either "important" or "non-important" modules. The corresponding confusion matrix is shown in Fig. 4, in which

- TP refers to the number of actual important modules that are predicted as important
- FP refers to the number of actual non-important modules that are predicted as important
- TN refers to the number of actual non-important modules that are predicted as non-important
- FN refers to the number of actual important modules that are predicted as non-important.

Based on the confusion matrix, the following indicators are used to evaluate the effectiveness of a prediction approach $m$:

$$\text{precision} = \frac{TP}{TP + FP}$$

$$recall = \frac{TP}{TP + FN}$$

---

[4] http://www.scitools.com

$$F_1 = \frac{2 \times precision \times recall}{precision + recall}$$

Each of the above indicators has a range between 0 and 1. In particular, a larger value indicates a better prediction effectiveness.

In practice, for a target project, it is very likely that only a very few modules are the truly important modules. Consequently, the prediction approach *m* may have a low precision even if the recall is high. If developers document all the predicted important modules, many falsely important modules will also be documented. This is no problem if the time/economic resource is enough. However, if the time/economic resource is not enough, i.e. not all the predicted important modules can be documented, it is wise to filter out falsely important modules (before documenting modules) by inspecting the predicted important modules. In this case, there is a strong need to evaluate the cost-benefit of *m*, which is ignored in McBurney et al.'s study [22]. Given a prediction approach, how to evaluate its cost-benefit in helping identify the truly important modules from the predicted important modules? Without loss of generalization, for a target project, assume that there are *n* modules in total, of which *k* module are truly important. For a prediction approach *m*, assume that, of all the *x* predicted important modules, *y* modules are truly important. In order to evaluate the cost-benefit of the prediction approach *m*, we make the following assumptions

- All the *x* predicted important modules are inspected;
- The inspection is perfect, i.e. all the *y* truly important modules in the inspected *x* modules are found;
- For computing a benefit, we use the random selection model (modules to be inspected are selected from the *n* modules by chance) as a baseline of comparison.

As a result, when using the prediction approach *m*, the percentage of modules needed to be inspected is $x/n \times 100\%$, while the percentage of truly important modules found is $y/k \times 100\%$. Similar to [48, 49], we evaluate how effective the prediction model *m* is in reducing the effort for inspection compared to a random model that achieves the same recall of finding truly important modules. To this end, we use the following ER (effort reduction) indicator:

$$ER = \frac{effort(random) - effort(m)}{effort(random)}$$

Here, effort(*m*) = $x/n \times 100\%$ and effort(*random*) = $y/k \times 100\%$. In other words, we have:

$$ER = \frac{\frac{y}{k} - \frac{x}{n}}{\frac{y}{k}}$$

A positive ER indicates that the prediction approach *m* is superior to a random model. A higher ER indicates a better cost-benefit.

By the above-mentioned four indicators (i.e. precision, recall, $F_1$, and ER), we can comprehensively evaluate the effectiveness of a prediction approach in documentation effort prioritization. In practice, the ER indicator is especially important when there is no enough time/economic resource to document all the predicted important modules.

*E. Evaluation method*

For each class in the test data, both the PageRank and ANN approaches will output a value ranging between 0 and 1, which can be explained as the predicted importance. As a result, the classes in the test data can be ranked in descending order according to the predicted importance. Given such a ranking list, developers can determine a threshold k% to classify the classes into two groups: the top k% as "important" and the bottom (100-k)% as "non-important" (in practice, developers can flexibly select a threshold according to their workload). In this context, we can compute the precision, recall, $F_1$, and ER indicators for both the PageRank and ANN approaches to compare their effectiveness. Similar to [22], we vary k% from 5% to 50% with a step length of 5% to conduct a comprehensive comparison between the PageRank and ANN approaches. Generally speaking, if an approach has higher precision, recall, $F_1$, and ER values under small thresholds, it is better in documentation effort prioritization.

For each data, we use the 100-times out-of-sample bootstrap [43] technique to conduct the experiment. According to [44], the out-of-sample bootstrap is the least biased model validation technique in terms of both threshold-dependent and threshold-independent performance measures. Assuming the size of the data set is *N*, an out-of-sample bootstrap uses the sampling with replacement to extract *N* instances as the training data and use the un-sampled instances as the test data. In our experiment, we run out-of-sample bootstrap 100 times. The seed used to generate random numbers in out-of-sample bootstrap is set as follows: 0 at the beginning and 1 is added for each time.

- For the ANN approach, at each run, the SCM and VSM models are trained using the training data and tested on the test data. After that, we use the predicted value to rank classes in descending order. Then, the top k% classes are selected as the predicted important classes.
- For the PageRank approach, we compute the importance for each class based on the classes in the data set (rather than in the whole project) and their dependence relationships. For the old data sets, this means that only the scored classes in McBurney et al.'s study are used. For example, for the jexcelapi-2.6.12 data sets, only 50 classes (rather than 458 classes) and their dependence relationships are analyzed to build the edge-weighted inter-module dependence graph.

For both the PageRank and ANN approaches, the final effectiveness is the average of the results on 100-times out-of-sample bootstrap, regardless of whether the precision, recall, $F_1$, or ER is taken into account.

## V. EXPERIMENTAL RESULTS

In this section, we report in detail our experimental results on the effectiveness of our PageRank approach in document effort prioritization, especially when compared with the state-of-the-art ANN approaches.

x_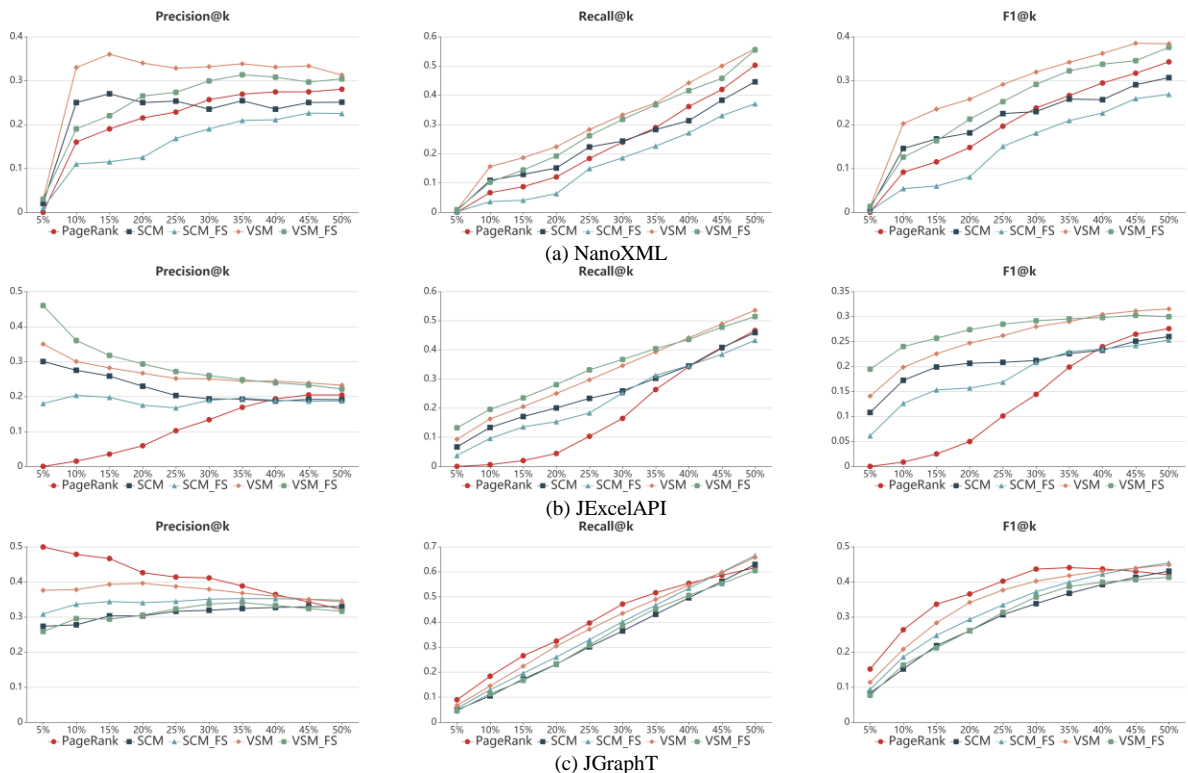

Fig. 5. Comparison of precision, recall and $F_1$ between the PageRank approach and the ANN approaches on the old data sets ("@k" denotes the corresponding results when k takes different thresholds)

*A. RQ1: How effective is the PageRank approach in identifying "important" classes compared with the state-of-the-art ANN approaches?*

In this section, we first give our two observations from the results, and give our analysis to each observation. More observations or analyses are put into RQ2 (a supplement to RQ1) for further elaboration and analysis.

**Observations**. From Fig. 5 (old data sets) and Fig. 6 (new data sets), we have the following observations: (1) compared with the ANN approaches, the PageRank approach performs very differently on the old data sets; (2) the PageRank approach is clearly superior to all the ANN approaches on the new data sets; and (3) with the decrease of threshold *k*, the precision (or $F_1$) of the PageRank approach increase more obviously than the ANN approaches on most projects.

Specifically, the first observation is as follows: On the JGraphT project, the PageRank approach is better than all the ANN approaches (SCM, SCM_FS, VSM, and VSM_FS). On the JExcelAPI project, the PageRank approach is inferior to all the ANN approaches. On the NanoXML project, the PageRank approach is between SCM (SCM_FS) and VSM (VSM_FS). The third observation is as follows: For the JGraphT project (Old data sets), the precision of the PageRank approach increases while all the ANN approaches decreases with the decrease of threshold. For all the projects in the new data sets, the precision and $F_1$ of the PageRank approaches increases more significantly than all the ANN approaches with the decrease of threshold.

**Analysis 1**. For the observation (1), we believe that there are two possible reasons for the large performance difference. First, the PageRank approach only uses a part of classes in the project to calculate the importance score, which may lose many dependence relationships among classes reflecting the importance of classes. Combining with Table I (a), it seems that the greater the difference between the number of classes used in the experiment and the total number of classes in a project, the worse the PageRank approach will be. We will discuss this point in section VI.A. Second, the total number of projects in the old data sets is too small (only 3) and the number of classes scored in McBurney et al.'s experiments is not large enough. It is likely that the old data set itself is not representative enough. Therefore, it is necessary to use more data sets to evaluate the effectiveness of the PageRank approach.

**Analysis 2**. For the observation (2), we believe that the main reason is that the PageRank approach successfully captures the essence of importance concept: the dependence relationships (especially the indirect dependences) among classes, rather than their characteristics such as size, complexity, or tf/idf, play a key role in determining which classes are important. For a given class, if many other classes directly or indirectly depend on it, it is naturally important for understanding the whole software system, no matter whether the class is simple or complex. For the ANN approaches, the used metrics include source code metrics (size, complexity, object-oriented metrics, and comment ratio) and tf/idf. Of these metrics, except few object-oriented metrics such as DIT and NSC, the remaining metrics have no explicit connection to the direct dependence relationships. Furthermore, all the source code and tf/idf metrics cannot capture the indirect dependence relationships (the PageRank approach tackles this problem by a recursive way to capture the indirect dependences). As a result, it is not surprised that the PageRank approach has an outstanding effectiveness in finding truly important classes compared with the state-of-the-art ANN approaches.

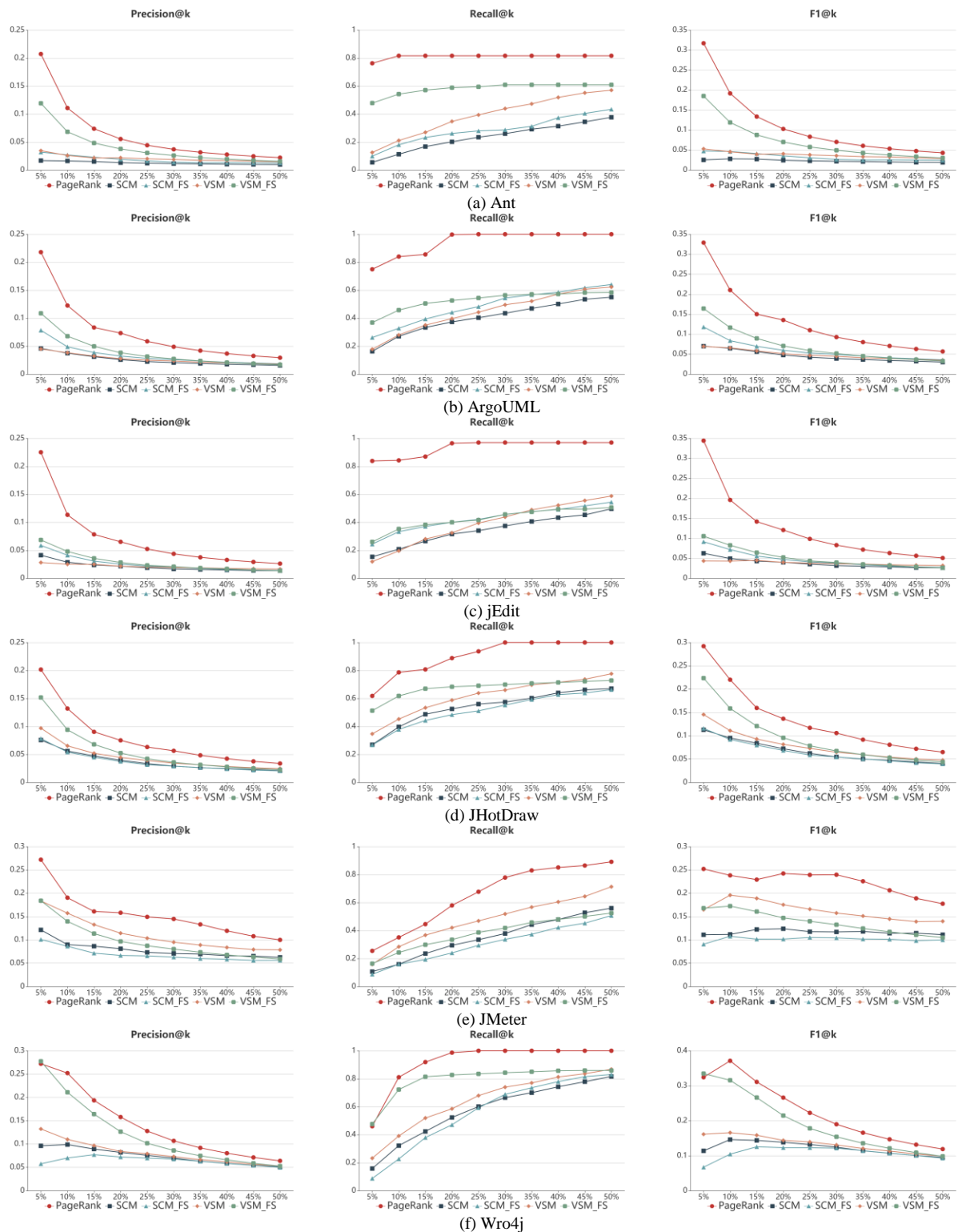

Fig. 6. Comparison of precision, recall and $F_1$ between the PageRank approach and the ANN approaches on the new data sets ("@k" denotes the corresponding results when k takes different thresholds).

**Analysis 3**. The results of observation (3) indicate that the PageRank approach gives higher importance scores to truly important classes, so the truly important classes rank higher (except for the NanoXML and JExcelAPI projects). Therefore, the PageRank approach works better at a smaller threshold (k% ≤ 25%). The main reason is that, under the PageRank approach, classes with rich inter-class dependencies are more likely to have larger importance scores. Intuitively, classes with rich dependencies can help programmers understand the functions of classes and the calling relationships of classes, so

TABEL II
Comparison of precision

(a) top 5%

| Project | precision | | | | |
|---|---|---|---|---|---|
| | SCM | SCM_FS | VSM | VSM_FS | PageRank |
| NanoXML | 0.02(N) | 0.01(N) | 0.03(N) | 0.03(N) | 0 |
| JExcelAPI | 0.30(S) | 0.18(S) | 0.35(M) | 0.46(M) | 0 |
| JGraphT | 0.27(M) | 0.31(M) | 0.38(S) | 0.26(L) | 0.5 |
| Ant | 0.02(L) | 0.03(L) | 0.03(L) | 0.12(M) | 0.21 |
| ArgoUML | 0.05(L) | 0.08(L) | 0.04(L) | 0.11(L) | 0.22 |
| jEdit | 0.04(L) | 0.06(L) | 0.03(L) | 0.07(L) | 0.23 |
| JHotDraw | 0.08(L) | 0.08(L) | 0.10(L) | 0.15(S) | 0.2 |
| JMeter | 0.12(M) | 0.10(L) | 0.18(S) | 0.18(S) | 0.27 |
| Wro4j | 0.10(L) | 0.06(L) | 0.13(L) | 0.28(N) | 0.27 |

(b) top 10%

| Project | precision | | | | |
|---|---|---|---|---|---|
| | SCM | SCM_FS | VSM | VSM_FS | PageRank |
| NanoXML | 0.25(N) | 0.11(N) | 0.33(S) | 0.19(N) | 0.16 |
| JExcelAPI | 0.28(M) | 0.20(M) | 0.30(L) | 0.36(L) | 0.02 |
| JGraphT | 0.28(L) | 0.34(M) | 0.38(S) | 0.30(L) | 0.48 |
| Ant | 0.02(L) | 0.03(L) | 0.03(L) | 0.07(M) | 0.11 |
| ArgoUML | 0.04(L) | 0.05(L) | 0.04(L) | 0.07(L) | 0.12 |
| jEdit | 0.03(L) | 0.04(L) | 0.03(L) | 0.05(L) | 0.11 |
| JHotDraw | 0.06(L) | 0.05(L) | 0.07(L) | 0.09(M) | 0.13 |
| JMeter | 0.09(L) | 0.09(L) | 0.16(S) | 0.14(S) | 0.19 |
| Wro4j | 0.10(L) | 0.07(L) | 0.11(L) | 0.21(S) | 0.25 |

TABEL III
Comparison of recall

(a) top 5%

| Project | recall | | | | |
|---|---|---|---|---|---|
| | SCM | SCM_FS | VSM | VSM_FS | PageRank |
| NanoXML | 0.01(N) | 0(N) | 0.01(N) | 0.01(N) | 0 |
| JExcelAPI | 0.07(S) | 0.04(S) | 0.09(M) | 0.13(M) | 0 |
| JGraphT | 0.05(M) | 0.06(M) | 0.07(S) | 0.05(L) | 0.09 |
| Ant | 0.05(L) | 0.10(L) | 0.13(L) | 0.48(M) | 0.76 |
| ArgoUML | 0.16(L) | 0.26(L) | 0.18(L) | 0.37(L) | 0.75 |
| jEdit | 0.15(L) | 0.24(L) | 0.12(L) | 0.26(L) | 0.84 |
| JHotDraw | 0.27(L) | 0.27(L) | 0.35(L) | 0.51(S) | 0.62 |
| JMeter | 0.11(M) | 0.09(L) | 0.16(S) | 0.16(S) | 0.25 |
| Wro4j | 0.16(L) | 0.09(L) | 0.23(L) | 0.48(N) | 0.46 |

(b) top 10%

| Project | recall | | | | |
|---|---|---|---|---|---|
| | SCM | SCM_FS | VSM | VSM_FS | PageRank |
| NanoXML | 0.11(N) | 0.04(N) | 0.16(S) | 0.10(N) | 0.07 |
| JExcelAPI | 0.13(L) | 0.10(M) | 0.16(L) | 0.20(L) | 0.01 |
| JGraphT | 0.11(L) | 0.13(M) | 0.14(M) | 0.11(L) | 0.18 |
| Ant | 0.11(L) | 0.18(L) | 0.21(L) | 0.54(M) | 0.82 |
| ArgoUML | 0.27(L) | 0.33(L) | 0.28(L) | 0.46(L) | 0.84 |
| jEdit | 0.21(L) | 0.33(L) | 0.20(L) | 0.35(L) | 0.84 |
| JHotDraw | 0.40(L) | 0.38(L) | 0.45(L) | 0.62(S) | 0.79 |
| JMeter | 0.16(L) | 0.16(L) | 0.28(S) | 0.24(S) | 0.35 |
| Wro4j | 0.32(L) | 0.23(L) | 0.39(L) | 0.72(S) | 0.81 |

TABEL IV
Comparison of F1

(a) top 5%

| Project | F1 | | | | |
|---|---|---|---|---|---|
| | SCM | SCM_FS | VSM | VSM_FS | PageRank |
| NanoXML | 0.01(N) | 0(N) | 0.01(N) | 0.01(N) | 0 |
| JExcelAPI | 0.11(S) | 0.06(S) | 0.14(M) | 0.19(M) | 0 |
| JGraphT | 0.08(M) | 0.09(M) | 0.11(S) | 0.08(L) | 0.15 |
| Ant | 0.03(L) | 0.05(L) | 0.05(L) | 0.19(L) | 0.32 |
| ArgoUML | 0.07(L) | 0.12(L) | 0.07(L) | 0.16(L) | 0.33 |
| jEdit | 0.06(L) | 0.09(L) | 0.04(L) | 0.11(L) | 0.34 |
| JHotDraw | 0.11(L) | 0.12(L) | 0.15(L) | 0.22(S) | 0.29 |
| JMeter | 0.11(L) | 0.09(L) | 0.16(S) | 0.17(S) | 0.25 |
| Wro4j | 0.11(L) | 0.07(L) | 0.16(L) | 0.33(N) | 0.32 |

(b) top 10%

| Project | F1 | | | | |
|---|---|---|---|---|---|
| | SCM | SCM_FS | VSM | VSM_FS | PageRank |
| NanoXML | 0.15(N) | 0.05(N) | 0.20(S) | 0.13(N) | 0.09 |
| JExcelAPI | 0.17(L) | 0.13(M) | 0.20(L) | 0.24(L) | 0.01 |
| JGraphT | 0.15(L) | 0.19(M) | 0.21(M) | 0.16(L) | 0.26 |
| Ant | 0.03(L) | 0.05(L) | 0.05(L) | 0.12(L) | 0.19 |
| ArgoUML | 0.06(L) | 0.08(L) | 0.07(L) | 0.12(L) | 0.21 |
| jEdit | 0.05(L) | 0.07(L) | 0.04(L) | 0.08(L) | 0.2 |
| JHotDraw | 0.10(L) | 0.09(L) | 0.11(L) | 0.16(M) | 0.22 |
| JMeter | 0.11(L) | 0.11(L) | 0.20(S) | 0.17(S) | 0.24 |
| Wro4j | 0.15(L) | 0.10(L) | 0.17(L) | 0.32(S) | 0.37 |

TABEL V
Comparison of ER

(a) top 5%

| Project | ER | | | | |
|---|---|---|---|---|---|
| | SCM | SCM_FS | VSM | VSM_FS | PageRank |
| NanoXML | 0.01(N) | 0(N) | 0.02(N) | 0.02(N) | 0 |
| JExcelAPI | 0.22(S) | 0.13(S) | 0.25(M) | 0.33(M) | 0 |
| JGraphT | 0.19(M) | 0.21(M) | 0.30(S) | 0.17(L) | 0.4 |
| Ant | 0.15(L) | 0.25(L) | 0.29(L) | 0.67(M) | 0.87 |
| ArgoUML | 0.40(L) | 0.48(L) | 0.45(L) | 0.67(L) | 0.92 |
| jEdit | 0.27(L) | 0.40(L) | 0.21(L) | 0.43(L) | 0.89 |
| JHotDraw | 0.40(L) | 0.41(L) | 0.60(M) | 0.75(S) | 0.85 |
| JMeter | 0.32(M) | 0.27(L) | 0.46(S) | 0.46(S) | 0.62 |
| Wro4j | 0.39(L) | 0.27(L) | 0.52(L) | 0.75(N) | 0.81 |

(a) top 10%

| Project | ER | | | | |
|---|---|---|---|---|---|
| | SCM | SCM_FS | VSM | VSM_FS | PageRank |
| NanoXML | 0.13(N) | 0.06(N) | 0.17(N) | 0.09(N) | 0.09 |
| JExcelAPI | 0.26(L) | 0.19(M) | 0.30(L) | 0.35(L) | 0.01 |
| JGraphT | -0.05(L) | 0.07(M) | 0.17(S) | -0.06(L) | 0.33 |
| Ant | 0.18(L) | 0.28(L) | 0.34(L) | 0.64(M) | 0.83 |
| ArgoUML | 0.42(L) | 0.47(L) | 0.47(L) | 0.65(L) | 0.87 |
| jEdit | 0.30(L) | 0.44(L) | 0.30(L) | 0.49(L) | 0.84 |
| JHotDraw | 0.47(L) | 0.47(L) | 0.60(L) | 0.72(S) | 0.83 |
| JMeter | 0.32(L) | 0.32(L) | 0.52(S) | 0.46(S) | 0.61 |
| Wro4j | 0.53(L) | 0.41(L) | 0.59(L) | 0.81(S) | 0.86 |

these classes may have a high degree of relevance to program understanding. Indeed, as analyzed in [39], classes that are truly important for program understanding are often rich in inter-class dependencies. Consequently, truly important classes tend to be ranked higher under the PageRank approach. For the ANN approaches, classes with larger size, complexity, or tf/idf metrics may have larger importance scores and hence tend to be ranked higher. However, these classes are not necessarily important for program understanding. This is the possible reason why the PageRank approach performs more outstandingly for a smaller threshold k% compared with the ANN approaches. Note that similar phenomena cannot be observed on the NanoXML and JExcelAPI projects. This may be due to the fact that a considerable proportion of classes and their related dependences in these two projects are not taken into account when the PageRank approach computes the importance scores for classes.

*B. RQ2: When only low thresholds are considered, how effective is the PageRank approach in identifying "important" classes compared with the state-of-the-art ANN approaches?*

RQ2 is a supplement to RQ1. In RQ1, the setting of threshold (the top 5%~50% classes are predicted as "important") is consistent with that of McBurney et al.' study [22]. In practice, it might make more sense to observe the effectiveness under smaller thresholds.

Tables II-V report the results, including precision, recall, F1, and ER, for the PageRank and ANN approaches under two

specific small thresholds (i.e. the top 5% and the top 10%). In each table, for each project, we first compare the PageRank approach against each of the ANN approaches to obtain the following two values: the p-value via the Wilcoxon signed-rank test [50] and the effect size via Cliff's δ [51]. Then, in order to control false discovery, we apply Benjamini-Hochberg's method [52] to adjust the above p-values. Finally, we mark the background of cells under the ANN approaches with different colors. A pink background means that the corresponding ANN approach is significantly worse than the PageRank approach (the adjusted p-value < 0.05). A blue background means that the corresponding ANN approach is significantly better (the adjusted p-value < 0.05). A white background means that the corresponding ANN approach is not significantly different from the PageRank approach (the adjusted p-value ≥ 0.05). In particular, we use "N", "S", "M", and "L" to respectively represent that the corresponding effect size is negligible ($|\delta| < 0.147$), small ($0.147 \le |\delta| < 0.33$), moderate ($0.133 \le |\delta| < 0.474$) and large ($|\delta| \ge 0.474$).

**Observations**. From Table II-V, we have the following observations: (1) except on the NanoXML and JExcelAPI projects, the PageRank approach is significantly better than the ANN approaches in terms of precision, recall, and $F_1$, with a large effect size in most cases; (2) at the thresholds of 5% and 10%, both PageRank approach and the ANN approaches have relatively low precision values (compared with other evaluation indicators); and (3) except on the NanoXML and JExcelAPI projects, the PageRank approach is much better than random method and the ANN approaches according to the cost-benefit ER, with a large effect size in most cases. At the thresholds of 5% and 10%, the ER values of PageRank on all indicators are higher than 0.6, and most of them are higher than 0.8. However, the ER values of the ANN approaches are all not higher than 0.6 except the VSM_FS approach. The VSM_FS approach, which has the highest ER value among the ANN approaches, even has a negative ER value on one project.

**Analysis 1**. The results of observation (1) further indicate that the PageRank approach gives higher importance scores to truly important classes, so the truly important classes rank higher. Based on the performance on the old and new data sets, we can see that the effectiveness and scalability of the ANN approaches are not as good as the PageRank approach. On the NanoXML project, the PageRank approach performs similarly to the ANN approaches. On the JExcelAPI project, the performance of the PageRank approach is worse than the ANN approaches. We believe that the possible reason is that the number of classes (i.e. the labeled classes) used for experiments is far less than the total number of classes in this project, which will lead to the loss of the dependencies among classes. As a result, the performance of the PageRank approach is suppressed. We will further discuss this point in section VI.A.

**Analysis 2**. For the observation (2), we believe that the main reasons are two-folds. First, on the new data set, the number of labeled important classes accounts for a relatively low proportion of the total number of classes (e.g. for Ant, the number of important classes (8) accounts for 1.2% of the total number of classes (664)), while the threshold (5% or 10%) is relatively high. This results in the precision not being very high. Second, in the old data set, although there are 25% important classes in each project, these important classes are based on the scores of several students rather than the scores of project developers or official documents of the project. In particular, according to [22], the students had not finished scoring all the classes in each project. Therefore, these labeled "important" classes may not be the truly important classes in the project for program comprehension, and these labeled "important" classes may also miss the truly important classes. This may lead to a reduction in precision.

According to the actual needs, moderately lowering the threshold may further improve the precision. However, we think that sometimes low precision is acceptable depending on the scenario and purpose. For example, in practice, recall may be more important in order to ensure the comprehension of team for the project. That is the most important classes needed to be documented as many as possible. At this point, a reduction in precision is tolerated, that is, some less important classes are also misidentified as important classes and are documented. At this point, there may be no loss to the comprehension of the project, but there may be some promotion significance.

**Analysis 3**. The results of observation (3) indicate that selecting important classes by the PageRank approach can indeed reduce the documentation effort developers need to write for program comprehension compared with the random approach and the ANN approaches. That is, the PageRank approach does achieve the prioritization of documentation effort. It is not difficult to find that with the threshold rising, the ER value of ANN approaches may be negative. This suggests that from the cost-benefit (ER) point of view, the ANN approaches may even increase the documentation effort. Therefore, the PageRank approach is more competent for the prioritization of documentation effort.

*The core observations based on the comparison results in RQ1 and RQ2 are as follows: the PageRank approach is superior to the-state-of-the-art supervised ANN approach on all indicators. The effectiveness and scalability of the-state-of-the-art supervised ANN approach are not as good as the PageRank approach. What is more, from the viewpoint of cost-benefit (ER), the PageRank approach is more competent for the prioritization of documentation effort than random approach and the ANN approach.*

## VI. DISCUSSIONS

In this section, we analyze the influence of various factors on the effectiveness of the PageRank approach. First, for the old data sets, we analyze the utility of classes in the whole project. Then, we analyze the influence of the weights of dependence relationships among classes.

### A. *The utility of classes in the whole project*

This section discusses the impact of the number of classes used for the experiment in the old data set on the PageRank approach. In the result of Fig. 5 in section V.A, our analysis reveals that the effectives of the PageRank approach on the old data sets is affected by the number of classes used for experiments. In the old data sets, the proportions of classes scored by graduate students in the three projects are as follows: NanoXML (75%), JExcelAPI (11%), and JGraphT (91%). As can be seen, none of these three projects has scored all the classes in each project. Because the PageRank approach uses the dependence relationships among classes to evaluate the importance of classes, it is natural that many dependence relationships are missing when only a part of classes are used. Therefore, the true performance of the PageRank approach is suppressed, especially for the JExcelAPI project.

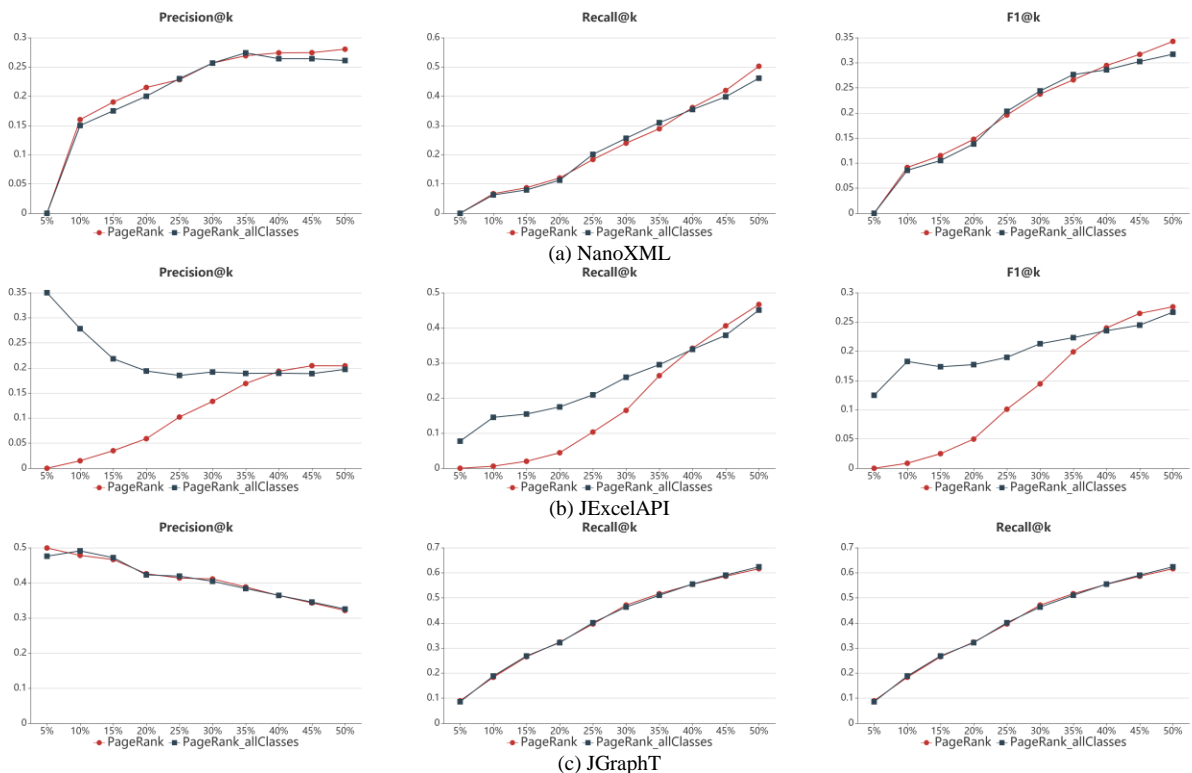

Fig. 7. Comparison of precision, recall, and $F_1$ between two types of PageRank approaches on the old data sets

To observe the influence of the number of classes, we design a new PageRank approach named "PageRank_allClasses", which uses all classes in a project to construct the dependence graph and recalculate the importance score of each class. Taking the JExcelAPI project as an example, the difference between the original PageRank approach in RQ1 and the PageRank_allClasses approach is that: the original PageRank approach uses 50 classes scored by graduate students to calculate the importance scores of each classes and then obtains their rankings. In contrast, the PageRank_allClasses approach first uses all the classes (i.e., 458) to calculate the importance scores of each classes and then extracts the classes that have actual category labels (i.e. 50 classes scored by users) to observe their rankings. The experimental settings and steps are the same as those in Section VI.

Fig. 7 shows a comparison of precision, recall, and $F_1$ between the original PageRank approach and the PageRank_allClasses approach on the old data sets. It can be observed that these two PageRank approaches have little difference on the NanoXML and JGraphT project. This is expected as the number of classes scored by graduate students in McBurney et al.'s study [22] is close to the total number of classes in each project. However, on the JExcelAPI project (which has low proportion of classes scored by graduate students in McBurney et al.'s study [22]), the effectiveness of the PageRank_allClasses approach has been greatly improved compared with the original PageRank approach.

The above result indicates that the number of classes used for the experiment in a project would affect the performance of the PageRank approach. The more classes are missing, the greater the influence on the PageRank approach will be. This is the reason why the original PageRank approach does not perform well on the JExcelAPI project (as shown in Section V RQ1 and RQ2). As a result, we suggest that if a future documentation effort prioritization study employs users (such as graduate students) to score the importance of a project, all the classes in the project should be scored. Otherwise, due to the insufficient representativeness of the scored classes, the resulting conclusion based on them may be biased.

B. *The influence of weights of dependence relationships*

As shown in equation (2) in Section III.C, for each class on a given project, the resulting *PR* (i.e. class importance) depends on $L_u$ and $TL(v)$. Since $L_u$ and $TL(v)$ are based on the edge weight in the inter-module class dependence graph, we know that, in nature, they depend on the weight of the four types of dependences: CI, CA, CM, and MM. As shown in Section III, in the PageRank approach, we assign the same weight to these four types of dependences. In other words, for the simplicity of computation, we do not distinguish their contributions when computing class importance.

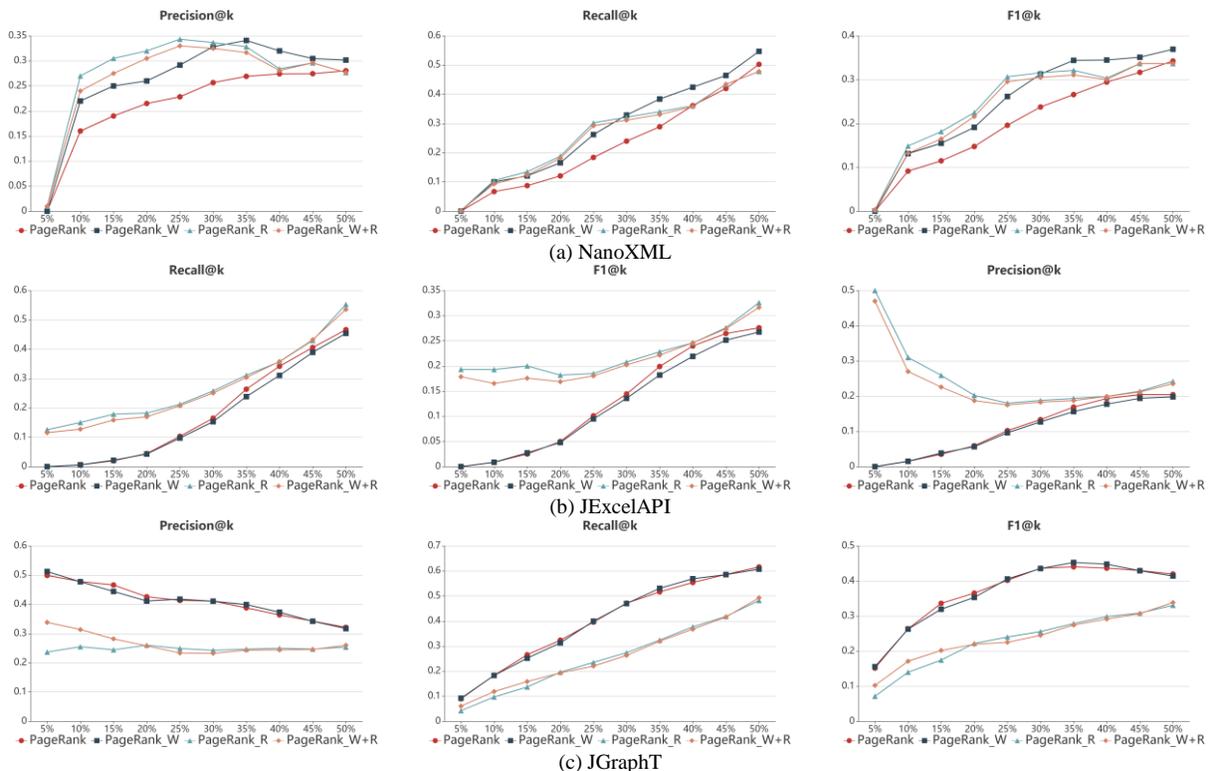

Fig. 8. Comparison of precision, recall, and $F_1$ between the original PageRank and three improved PageRank approaches on the old data sets

In the following, we investigate the influence of weights of dependence relationships. Following the work in [39], we use the following two methods to assign the weights to dependence relationships:

- Empirical weight. Literature [39] believed that different dependences had different contributions when computing class importance and hence assigned different multiplication coefficients to different dependences when expressing the weight of edges. We call this improved method of assigning weights empirical weighting. Referring to the settings in literature [39], we assign the multiplication coefficient (3, 3, 2, and 4) to the four types of dependence relationships (CI, CA, CM and MM) used in this paper. In this way, the equation of weight

    W(u, v) = CI(u, v)+CA(u, v)+CM(u, v)+MM(u, v)

    is changed to

    W(u, v) = 3CI(u, v)+3CA(u, v)+2CM(u, v)+4MM(u, v).

- Back recommendation. In literature [39], they call the edge from A to B a forward recommendation and the edge from B to A a back recommendation. In particular, "the weight of the forward recommendation from A to B is given by the dependency strength of the cumulated dependencies from A to B. The weight of the back recommendation from B to A is a fraction F of the weight of the forward recommendation from A to B" [39]. Let the weight matrix of the class dependence graph only using forward recommendation be $R$, then the weight matrix of the class dependence graph adding back recommendation be $R + \frac{1}{F} \times R^T$. Here, $T$ represents the matrix transpose.

The class dependence graph corresponding to equation (2) has only "forward recommendation" edge, i.e. the weight matrix is $R$. (Note: the definition of forward recommendation edge in [39] is the same as the Out-Edge in Section III.B). Therefore, when we combine "backward recommendation" to improve the weight matrix $R$, $R$ is changed to $R + \frac{1}{2} \times R^T$, where $\frac{1}{2}$ is the best value of $\frac{1}{F}$ reported in literature [39].

For the simplicity of presentation, we use "PageRank_W" to denote the PageRank approach using empirical weights, use "PageRank_R" to denote the PageRank approach using back recommendation, and use "PageRank_W+R" to denote the PageRank approach using both.

We repeat the experimental steps in section V to obtain the results of "PageRank_W", "PageRank_R", and "PageRank_W+R". Fig. 8 and 9 show the comparisons of precision, recall, and $F_1$ among these four PageRank approaches. From these figures, we make the following observations. First, PageRank_W is close to PageRank and PageRank_W+R is close to PageRank_R on almost all projects (except PageRank_W is evidently not close to PageRank on the NanoXML project). This shows that empirical weights have less impact on the PageRank model than back recommendation. Second, compared with the original PageRank approach, PageRank_R and PageRank_W+R, which use the back recommendation, show completely different effectiveness on different projects: some are obviously improved (e.g. NanoXML, JExcelAPI, JHotDraw, and JMeter), some are obviously decreased (e.g. JGraphT and Ant), and some are little changed (e.g. ArgoUML, jEdit, and Wro4j).

Therefore, for the PageRank approach used in this paper, the influence of empirical weights is relatively small, and the influence of back recommendation is relatively large. Since empirical weights and back recommendation do not always improve or decrease PageRank performance, we suggest that these two improvement methods for PageRank should be used cautiously in practice.

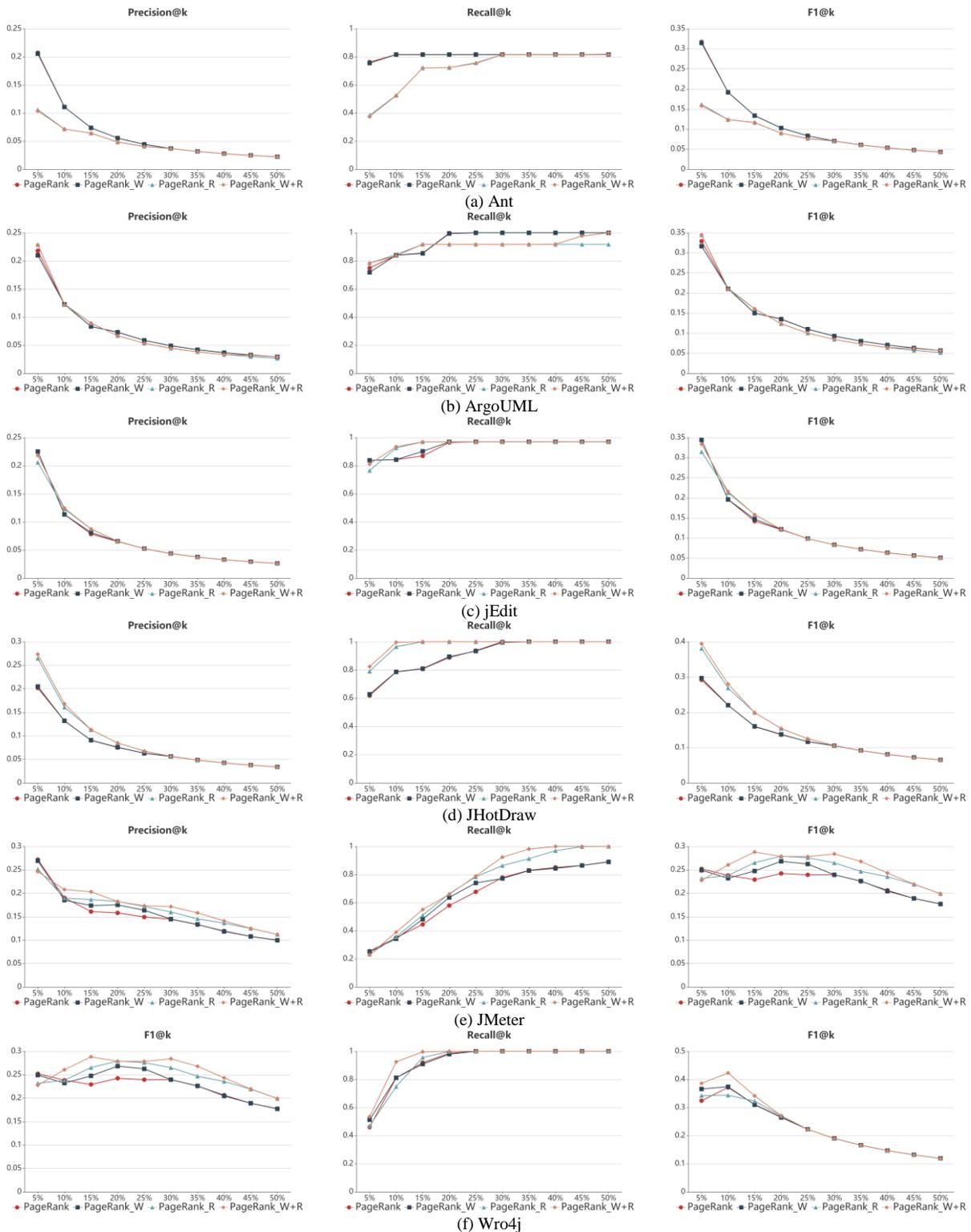

Fig. 9. Comparison of precision, recall, and $F_1$ between the original PageRank and three improved PageRank approaches on the new data sets

## VII. THREATS TO VALIDITY

In this section, we discuss the threats to validity of our study, including construct validity, internal validity, and external validity.

1616

## A. Construct validity

*Construct validity* is the degree to which the independent and dependent variables accurately measure the concept they purport to measure. In our study, the used dependent variable is a binary label that indicates whether a class is important. For the old data sets, we use the labels shared online by McBurney et al. [22]. These labels are set according to the importance scores they collected from graduate students. As mentioned in [22], for a given class, they used the average of the scores over methods in the class given by graduate students as its importance score. If a method was not scored, it was not considered when computing the corresponding class importance. Therefore, the labels on the old data sets may be inaccurate, either due to the subjective bias caused by graduate students or due to the missing scores of methods in a class. For the new data sets, according to the description in literature [39], they collected the labels via information from diverse sources, including the project tutorial, design description, and development documents (most from official documents). In this sense, the label formation on the new data sets should be accurate and hence the construct validity of the dependent variable should be considered satisfactory.

The independent variables used in this study are static code metrics, tf/idf metrics, and PageRank scores, each of which has a clear definition. For the static code metrics and tf/idf metrics, we use scripts on mature tools such as Understand and Weka to collect the data. For the PageRank scores, we developed our scripts to collect the data, which had been sufficiently tested and had been used in our previous studies for a long time. In this sense, the construct validity of the independent variable should be acceptable in our study.

## B. Internal validity

*Internal validity* is the degree to which conclusions can be drawn about the causal effect of independent variables on the dependent variable. There are two possible threats to the internal validity. The first threat is the unknown effect of the deviation of the independent variables from the normal distribution. In our study, we used the raw data to build the ANN models when investigating RQ1 and RQ2. In other words, we did not take into account whether the independent variables follow a normal distribution. The reason is that, in ANN, there is no assumption related to normal distribution. However, previous studies suggested applying the log transformation to the independent variables to make them close to a normal distribution, as it might lead to a better model [88]. To reduce this threat, we applied the log transformation and rerun the analyses. We found that the conclusions for RQ1 and RQ2 did not change before and after the log transformation. The second threat is the unknown effect of feature selection method. In our study, we use CFS as the feature select method when building the ANN models. In order to reduce this threat, we used other representative feature selection methods such as ReliefF, InfoGain, and GainRatio [90] to reran the analysis and found the results to be very similar.

## C. External validity

*External validity* is the degree to which the results of the research can be generalized to the population under study and other research settings. The most important threat is that our finding may not be generalized to non-Java projects or commercial projects. All projects in both the old data sets [22] and new data sets [39] used in this paper are open-source Java projects. These projects have their own particularities and peculiarities, so they may not be generally representative. To mitigate this threat, there is a need to replicate our study across a wider variety of projects in the future work.

## VIII. CONCLUSION AND FUTURE WORK

The goal of prioritizing code documentation effort is to identify modules that are important to software quality assurance activities, and the documentation effort of these modules should take precedence. In this paper, we propose an unsupervised PageRank approach to prioritizing documentation effort. This approach identifies important modules only based on the dependence relationships between modules in a project. As a result, the PageRank approach does not need any training data to build the prediction model. Based on our experimental results on six new added large data sets and three data sets used in previous studies [22], we find that the PageRank approach is superior to the-state-of-the-art supervised ANN approach. In particular, the effectiveness and scalability of the ANN approach is not as good as the PageRank approach. What is more, from the cost-benefit point of view, PageRank is more competent for the prioritization of documentation effort than the random and the ANN approaches. Due to the simplicity and effectiveness, we strongly recommend that the PageRank approach should be used as an easy-to-implement baseline in future research on documentation effort prioritization, and that any new approach should be compared with the PageRank approach to demonstrate its effectiveness.

In future work, we plan to do more empirical studies on non-Java or closed-source projects to evaluate the effectiveness of the PageRank approach. In addition, we plan to conduct more theoretical analysis and experiments to explore the corresponding relationship between the number of modules in a project and the effectiveness of the PageRank approach. Meanwhile, we will try more strategies to further improve the PageRank approach.

APPENDIX A

Table VI
List of **S**tatic source **C**ode **M**etrics (SCM) and **T**extual **C**omparison **M**etrics (TCM) in McBurney et al.'s study [22]

| Type | Metric | Description | Tools for measuring metrics |
|---|---|---|---|
| SCM: Size | LOC | Number of lines of code including comments but not empty lines | |
| | Statements | Number of executable code statements | |
| SCM: Complexity | %Branch | Branch statements account for percentage of statements | SourceMonitor[1] |
| | Calls | Number of statements for method calls | |
| | Calls Per Statement | Number of statements for method calls / Statements | |
| | Methods Per Class | Average number of methods for each class | |
| | Statements Per Method | Average number of statements contained in each method | |
| | Avg. Depth | Average number of branch layers nested in a function | |
| | Max Depth | Maximum number of branch layers nested in a function | |
| | Avg. Complexity | Average McCabe Cyclomatic Complexity of methods | |
| | Max Complexity | Maximum McCabe Cyclomatic Complexity of methods | |
| | WMC | The sum of McCabe Cyclomatic Complexity of methods per class | Metrics[2] |
| | NOF | Number of fields of a class | |
| SCM: Object Oriented | DIT | Number of ancestor classes a given class has | Metrics[2] |
| | NSC | Number of children classes a given class has | |
| | LCOM | Lack of cohesion of methods | |
| | NORM | Number of methods in a class overridden by its child classes | |
| | Abstract | A class is or is not an abstract class | |
| SCM: Others | %Comments | Annotated line account for percentage of all lines | SourceMonitor[1] |
| TCM | Class Appearance | The class name appears or doesn't appear in the two bodies of text | Scripts[4] |
| | Package Appearance | The package name appears or doesn't appear in the two bodies of text | |
| | Combination Appearance | The class and package name appears or doesn't appear in the two text | |
| | First Overlap[3] metric | Overlap similarity that words with splitting on camel case. | |
| | Second Overlap[3] metric | Overlap similarity that words without splitting on camel case. | |
| | First STASIS[3] metric | STASIS similarity that words with splitting on camel case. | |
| | Second STASIS[3] metric | STASIS similarity that words without splitting on camel case. | |

1. **SourceMonitor** is a tool for collecting static source code metrics, which can be downloaded from: http://www.campwoodsw.com
2. **Metrics** is an Eclipse plugin to extract object-oriented metrics or complexity metrics.
3. **Overlap** and **STASIS** are textual and semantic similarity metrics. For specific definitions, please refer to the paper [22].
4. TCM are collected by McBurney et al.'s script [22], in which the script that collect the **STASIS** metric can be accessed from http://www.cis.upenn.edu/~paulmcb/research/doceffort/.

20[82] F. Long, M.C. Rinard. An analysis of the search spaces for generate and validate patch generation systems. ICSE 2016: 702-713.
[83] M. Martinez, T. Durieux, R. Sommerard, J. Xuan, M. Monperrus. Automatic repair of real bugs in java: a large-scale experiment on the defects4j dataset. Empirical Software Engineering, 22(4), 2017: 1936-1964.
[84] Y. Xiong, J. Wang, R. Yan, J. Zhang, S. Han, G. Huang, L. Zhang. Precise condition synthesis for program repair. ICSE 2017: 416-426.
[85] X.D. Le, D. Lo, C.L. Goues. History driven program repair. SANER 2016: 213-224.
[86] F. Long, M. Rinard. Automatic patch generation by learning correct code. POPL 2016: 298-312.
[87] S. Mechtaev, J. Yi, A. Roychoudhury. Angelix: scalable multiline program patch synthesis via symbolic analysis. ICSE 2016: 691-701.
[88] T. Menzies, J. Greenwald, A. Frank. Data mining static code attributes to learn defect predictors. IEEE Transactions on Software Engineering, 33(1), 2007: 2-13.
[89] Y. Zhou, Y. Yang, H. Lu, L. Chen, Y. Li, Y. Zhao, J. Qian, B. Xu. How far we have progressed in the journey? An examination of cross-project defect prediction. ACM Transactions on Software Engineering and Methodology, 27(1), article 1, 2018: 1-51.
[90] B. Ghotra, S. McIntosh, A.E. Hassan. A large-scale study of the impact of feature selection techniques on defect classification models. MSR 2017: 146-157.
20